\documentclass[aps,pra,preprint,preprintnumbers,amsmath,amssymb,floatfix,showpacs]{revtex4-2}
\usepackage[cp1251]{inputenc}
\usepackage{graphicx}
\usepackage{epsfig}
\newcommand{\eref}[1] {(\ref{#1})}
\newcommand{\Eref}[1] {Eq.~(\ref{#1})}
\newcommand{\Tref}[1] {Table \ref{#1}}
\newcommand{\Fref}[1] {Fig. \ref{#1}}
\newcommand{\np}{\newpage}
\newcommand{\bt}{\begin{tabular}}
\newcommand{\et}{\end{tabular}}
\newcommand{\mc}{\multicolumn}

\begin{document}

\title{Time-dependent convergent close coupling method for\\ molecular ionization in laser fields}

\author{Vladislav V. Serov} \affiliation{Department of General, Theoretical and Computer Physics, Saratov State University, 83 Astrakhanskaya, Saratov 410012, Russia}

\begin{abstract}
We develop a time-dependent multi-configurational numerical technique for calculating ionization by short laser pulses of many-electron molecules. 
The method is based on the expansion of the wave function of a molecule into the eigenstates of the molecular ion.
We classify this method as time-dependent convergent close coupling (TDCCC) because it uses the same symmetry for channel functions as the well-known convergent close coupling (CCC) method.
\end{abstract}

\pacs{34.80.Gs, 34.80.Dp, 33.80.Eh}

\maketitle

\newpage


\section{Introduction}

Apparently, the only approach that allows one to correctly describe
the nuclear and electronic dynamics during single ionization of a
molecule without enormous expenditure of computational resources is
the multi-configuration method.

To date, many variants of this approach have been proposed and
used. The most advanced method appears to be the one developed by the
group of F. Martin et al \cite{FMartinReview,FMartin2016}. This method is based on the use of
pseudostates obtained by diagonalizing the Hamiltonian on a hybrid
basis, assembled from functions of three types: Gaussian
multicenter molecular orbitals, single-center Gaussian functions, and
b-splines \cite{FMartin2016}. 
Multicenter Gaussian functions are used to approximate the wave function near the nuclear core. 
The advantages of using them are fast calculation of multicenter matrix elements and compatibility with programs for quantum chemical calculations that use the same basis. On the contrary, b-splines depending on the radial variable are used to approximate the far part of the wave function describing escaping electrons \cite{FMartin2001}. 


Nevertheless, despite all the advantages, this method also has
disadvantages. The size of matrices and, accordingly, the speed of
calculations for time evolution grows quadratically with the number of
pseudostates. So increasing the energy resolution, or the maximum
energy of emitted electrons, causes calculations to slow down
dramatically. And for problems like Reconstruction of Attosecond
Beating By Interference of Two-photon Transitions (RABBIT) with the
resolution of nuclear states \cite{Wang2021}  or the observation of autoionization
states, a high energy resolution is required.

Therefore, we developed a method that does not use pseudostates for
the continuum.

\section{TDCCC formalism}

Consider a two-electron molecule with the total electron spin $S$ and
fixed nuclei. Time-dependent Schr\"dinger equation for the coordinate
part of the wave function is
\begin{eqnarray}
i\frac{\partial\Psi}{\partial t}(\mathbf{r}_1, \mathbf{r}_2,t) =
\hat{H} \Psi(\mathbf{r}_1, \mathbf{r}_2,t) \label{TDSE2}
\end{eqnarray}
where the Hamiltonian can be written as
\begin{eqnarray}
\hat{H}= \hat{h}_0(\mathbf{r}_1)+
\hat{h}_0(\mathbf{r}_2)+v(\mathbf{r}_1,\mathbf{r}_2)+\hat{w}(\mathbf{r}_1,t)+\hat{w}(\mathbf{r}_2,t).
\label{H_2e}
\end{eqnarray}

Here $\hat{h}_0(\mathbf{r})$ is the Hamiltonian of a singly ionized
ion, $v(\mathbf{r}_1,\mathbf{r}_2)$ is the potential of
electron-electron interaction, $\hat{w}(\mathbf{r},t)$ is the
Hamiltonian of interaction with an external field.

Since we are interested in the ionization process, we will construct a
multi-configuration wave function based on the function
$\phi_n(\mathbf{r})$, that satisfy the stationary Schr\"dinger equation
for the ion
\begin{eqnarray}
\hat{h}_0 \phi_n(\mathbf{r}) = \epsilon_n \phi_n(\mathbf{r}). 
\end{eqnarray}
The multi-configuration wave function can be written as
\begin{eqnarray}
\Psi(\mathbf{r}_1, \mathbf{r}_2,t) = \sum_{n=1}^{N_s}
    [\psi_n(\mathbf{r}_1,t) \phi_n(\mathbf{r}_2) + (-1)^S
      \phi_n(\mathbf{r}_1) \psi_n(\mathbf{r}_2,t) ] ,
\label{Psi_2e} 
\end{eqnarray}
where $N_s$ is the number of ion states used. This structure of the
function guarantees that after the emission of one electron, 
the remaining electron will end up in one of the ion's own states.

If we project both sides of \Eref{TDSE2} onto $\phi_n(\mathbf{r}_2)$,
we obtain the equation
\begin{eqnarray}
i\frac{\partial \langle\phi_n(\mathbf{r}_2)|\Psi\rangle}{\partial t} =
\langle\phi_n(\mathbf{r}_2)|\hat{H}| \Psi \rangle
\end{eqnarray}
Substituting function \eref{Psi_2e} here, we obtain the equation
\begin{eqnarray}
&& i\frac{\partial\psi_n(\mathbf{r}_1,t)}{\partial t} + (-1)^S
  i\frac{d}{dt} \sum_{m=1}^{N_s}
  \langle\phi_n(\mathbf{r}_2)|\psi_m(\mathbf{r}_2,t)\rangle
  \phi_m(\mathbf{r}_1)
=  
\nonumber \\ && \sum_{m=1}^{N_s}
\langle\phi_n(\mathbf{r}_2)|\hat{H}|\phi_m(\mathbf{r}_2)\rangle
\psi_m(\mathbf{r}_1,t) + (-1)^S \sum_{m=1}^{N_s}
\langle\phi_n(\mathbf{r}_2)|\hat{H}|\psi_m(\mathbf{r}_2,t)\rangle
\phi_m(\mathbf{r}_1)
\label{evolEq_2e}
\end{eqnarray}
This equation is inconvenient because it contains two terms with time
derivatives. To solve this problem, let's introduce a new function
\begin{eqnarray}
\bar{\psi}_n(\mathbf{r}_1,t) = \psi_n(\mathbf{r}_1,t) + (-1)^S
\sum_{m=1}^{N_s}
\langle\phi_n(\mathbf{r}_2)|\psi_m(\mathbf{r}_2,t)\rangle
\phi_m(\mathbf{r}_1)
\end{eqnarray}

The introduction of a new function allows us to combine two terms with
time derivatives into one.

To simplify the expression on the right side, consider the properties
of the new function. The projections of the new function onto the
basis functions satisfy the relation
\begin{eqnarray*}
\langle\phi_k(\mathbf{r})|\bar{\psi}_n(\mathbf{r},t)\rangle
= \langle\phi_k(\mathbf{r})|\psi_n(\mathbf{r},t)\rangle
+ (-1)^S \langle\phi_n(\mathbf{r})|\psi_k(\mathbf{r},t)\rangle 
\end{eqnarray*}
from which it follows that it must satisfy the symmetry condition
\begin{eqnarray}
\langle\phi_n(\mathbf{r})|\bar{\psi}_k(\mathbf{r},t)\rangle = (-1)^S
\langle\phi_k(\mathbf{r})|\bar{\psi}_n(\mathbf{r},t)\rangle
\label{CCC_symmetry}
\end{eqnarray}
A similar condition was once proposed to impose the convergent
close-coupling \cite{IBray1992}, so in what follows we will call such
relations the CCC symmetry.

Similar to the time-independent CCC \cite{IBray1992}, the same
function $\Psi$ can be written in terms of different
$\psi_n(\mathbf{r}_1,t)$, so without loss of generality we can impose
the CCC symmetry condition on the original function, i.e. put that
$\langle\phi_n(\mathbf{r})|\psi_k(\mathbf{r},t)\rangle = (-1)^S
\langle\phi_k(\mathbf{r})|\psi_n(\mathbf{r},t)\rangle$.  Then
$\langle\phi_k(\mathbf{r})|\bar{\psi}_n(\mathbf{r},t)\rangle = 2
\langle\phi_k(\mathbf{r})|\psi_n(\mathbf{r},t)\rangle$, and we can
express the old function via the new one
\begin{eqnarray}
\psi_n(\mathbf{r}_1,t) &=& \bar{\psi}_n(\mathbf{r}_1,t) - \frac{1}{2}
\sum_{m=1}^{N_s}
\langle\phi_m(\mathbf{r}_2)|\bar{\psi}_n(\mathbf{r}_2,t)\rangle
\phi_m(\mathbf{r}_1)
\end{eqnarray}
Let us introduce another auxiliary function obtained by
orthogonalization of the function with respect to the basis functions
\begin{eqnarray}
\tilde{\psi}_n(\mathbf{r},t) = \bar{\psi}_n(\mathbf{r},t) - 
\sum_{k=1}^{N_s} \langle\phi_k|\bar{\psi}_n\rangle \phi_k(\mathbf{r}) 
\label{tilde_psi}
\end{eqnarray}
With the use of the newly introduced functions and the CCC symmetry,
we can rewrite \Eref{evolEq_2e} in the form
\begin{eqnarray}
i\frac{\partial\bar{\psi}_n(\mathbf{r}_1,t)}{\partial t} &=&
[\hat{h}(\mathbf{r}_1,t)+\epsilon_n]\bar{\psi}_n(\mathbf{r}_1,t) +
\sum_{m=1}^{N_s} \hat{V}_{nm} \bar{\psi}_m(\mathbf{r}_1,t) +
\sum_{m=1}^{N_s} \hat{X}_{nm} \tilde{\psi}_m(\mathbf{r}_1,t)
\label{evolEqVX_2e}
\end{eqnarray}
where
$\hat{h}(\mathbf{r},t)=\hat{h}_0(\mathbf{r})+\hat{w}(\mathbf{r},t)$. 
In this equation, the explicit exchange term (the last term on the
right) describes only the exchange with the ionic states that are not
included in the basis, and the exchange with the basis states is
hidden in the remaining terms due to the use of the CCC symmetry.

Here we introduce the matrix of inter-channel potentials
\begin{eqnarray}
\hat{V}_{nm}(\mathbf{r}_1,t) = \langle\phi_n(\mathbf{r}_2)|v(\mathbf{r}_1,\mathbf{r}_2)|\phi_m(\mathbf{r}_2)\rangle + \langle\phi_n(\mathbf{r}_2)|\hat{w}(\mathbf{r}_2,t)|\phi_m(\mathbf{r}_2)\rangle
\end{eqnarray}
The first term on the right depends only on $\mathbf{r}_1$, the second
- only on $t$.

We define the exchange operator for an arbitrary function as
\begin{eqnarray}
\hat{X}_{nm} \psi &=&
\langle\phi_n(\mathbf{r}_2)|\hat{H}|\psi(\mathbf{r}_2,t)\rangle
\phi_m(\mathbf{r}_1) \label{Xoper}
\end{eqnarray}
For the function $\tilde{\psi}_m(\mathbf{r}_2,t)$, due to its
orthogonality to the ionic basis states, only terms with
inter-electronic and external potentials remain in the exchange
operator
\begin{eqnarray}
\hat{X}_{nm} \tilde{\psi}_m &=& [\langle\phi_n(\mathbf{r}_2)|v(\mathbf{r}_1,\mathbf{r}_2)|\tilde{\psi}_m(\mathbf{r}_2,t)\rangle + \langle\phi_n(\mathbf{r}_2)|\hat{w}(\mathbf{r}_2,t)|\tilde{\psi}_m(\mathbf{r}_2,t)\rangle]
\phi_m(\mathbf{r}_1) 
\label{Xoper_tilde}
\end{eqnarray}

The CCC symmetrization procedure can be described as the operator
$\hat{C}$, the action of which on the function is expressed as
\begin{eqnarray}
\hat{C}\psi_n(\mathbf{r}_1,t)= \frac{1}{2}\sum_{m, n',m'}
    [\delta_{nn'}\delta_{mm'}+(-1)^S\delta_{mn'}\delta_{nm'}]
    \langle\phi_{m'}(\mathbf{r}_2)|\bar{\psi}_{n'}(\mathbf{r}_2,t)\rangle
    \phi_m(\mathbf{r}_1) + \tilde{\psi}_n(\mathbf{r}_1,t)
\label{CCCsymmetrization} 
\end{eqnarray}

\section{Numerical implementation}

To represent the radial dependence of the wave function, discrete
variable representation (DVR) was used based on the finite elements of
the Gauss-Lobatto quadrature. The angular dependence was represented
by a spherical harmonic expansion \cite{Serov2011}.

The time evolution in \Eref{evolEqVX_2e} was carried out using an
implicit fourth-order scheme \cite{Selin1999}, which is a
generalization of the Crank-Nicholson scheme. The system of linear
equations to which this scheme leads was solved using the biconjugate
gradient method with a preconditioner. Each multiplication by the
Hamiltonian matrix was accompanied by the CCC symmetrization
\eref{CCCsymmetrization} before and after the multiplication, which
ensured the CCC symmetry of the function.

Suppression of the unphysical reflection from the radial grid
boundary was carried out using exterior complex scaling (ECS) for the
radial variable.

Ionization amplitudes were extracted from the calculated wave function
using the t-SURFFc method \cite{Serov2013}.

\np
\section{Test results}

\begin{table}[h]
\caption{Energy (in atomic units a.u.) of various two-electron
  stationary states.}
\label{tab:Convergence}
\centering
\begin{tabular}{|c|c|c|c|c|}
\hline
$N_s$ & He 1s$^2\,{}^1$S & He 1s2s$\,{}^3$S & Li$^+$ 1s$^2\,{}^1$S & H$_2$ ($R=1.4$) \\
\hline
1          & -2.87203 &          & -7.24148 & -1.85875 \\
5          & -2.88415 & -2.17022 & -7.25358 & -1.86796 \\
14         & -2.88614 &          & -7.25583 &          \\
exact      & -2.90339 \cite{NIST} & -2.17503 \cite{NIST} & -7.27984 \cite{NIST} & -1.88876 \cite{Kolos1968} \\
\hline
\end{tabular}
\end{table}

In the first test, we check the convergence of the energy of the
ground states as the number of configurations $N_s$ in
\Eref{Psi_2e} increases. 
The energy of the ground states was calculated by evolution in imaginary time.
 As the two-electron targets, we consider the
He and Li atoms as well as the H$_2$ molecule. The atoms in the
present molecular formalism correspond to the limit of the
inter-nuclear distance set to zero.  Our test results in comparison
with the literature values are shown in \Tref{tab:Convergence}. 
It can be seen that adding more ionic states results in only a small change in the calculated ground state energies. This is a well-known effect for variational calculations: if the selected basis functions do not have the same set of singularities as the exact function, then the first few functions make the main contribution to the energy, but a lot of functions are needed to converge to the exact value. We can't use very many ionic states in calculations without greatly slowing down the program. But we are willing to pay a fee in the form of inaccuracies in the energy of stationary states to ensure the correctness of the wave function after ionization.

\begin{figure}[ht!]
	\centering
	\includegraphics[width=0.7\textwidth]{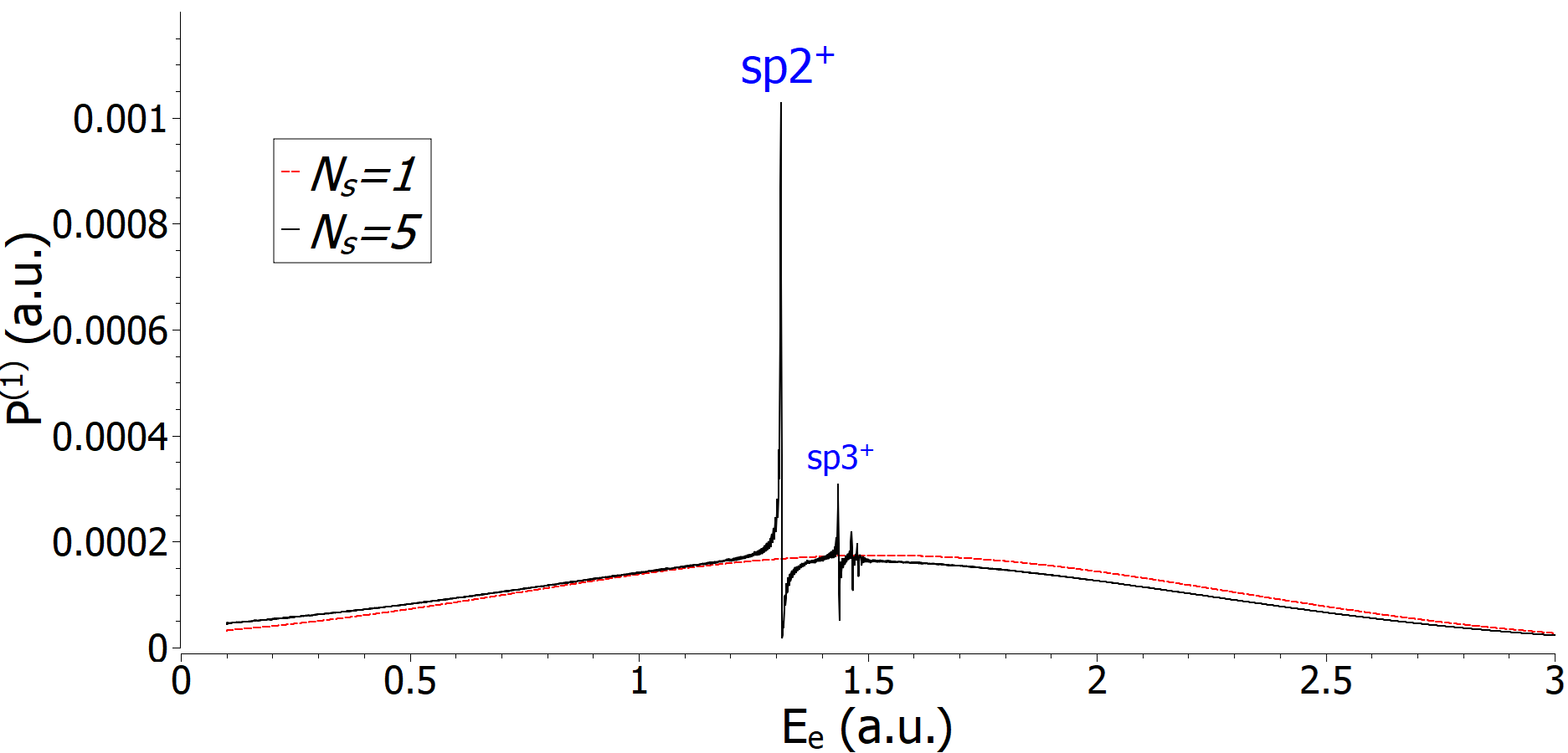}
	\includegraphics[width=0.7\textwidth]{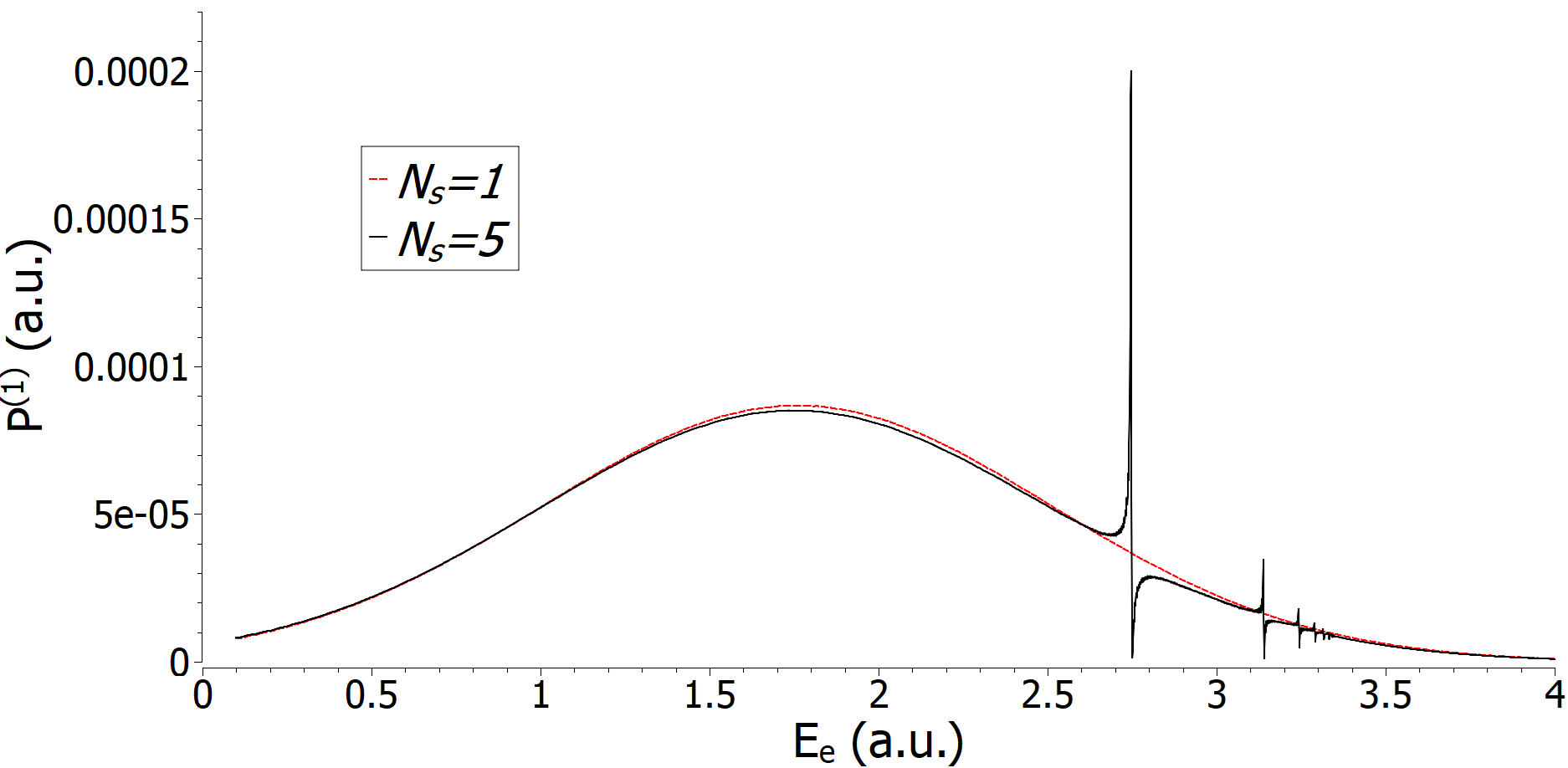}
	\caption{Spectrum of electrons emitted after ionization by a
          short XUV pulse with a FWHM=38 as in the cases $N_s=1$ (red
          dash line) and $N_s=5$ (black solid line): (top) He,
          $\omega_{XUV}=2.679$ a.u.; (bottom) Li$^+$,
          $\omega_{XUV}=4.6$ a.u.}
	\label{fig:SpectrumShortPulse}
\end{figure}

\begin{figure}[ht!]
	\centering
	\includegraphics[width=0.7\textwidth]{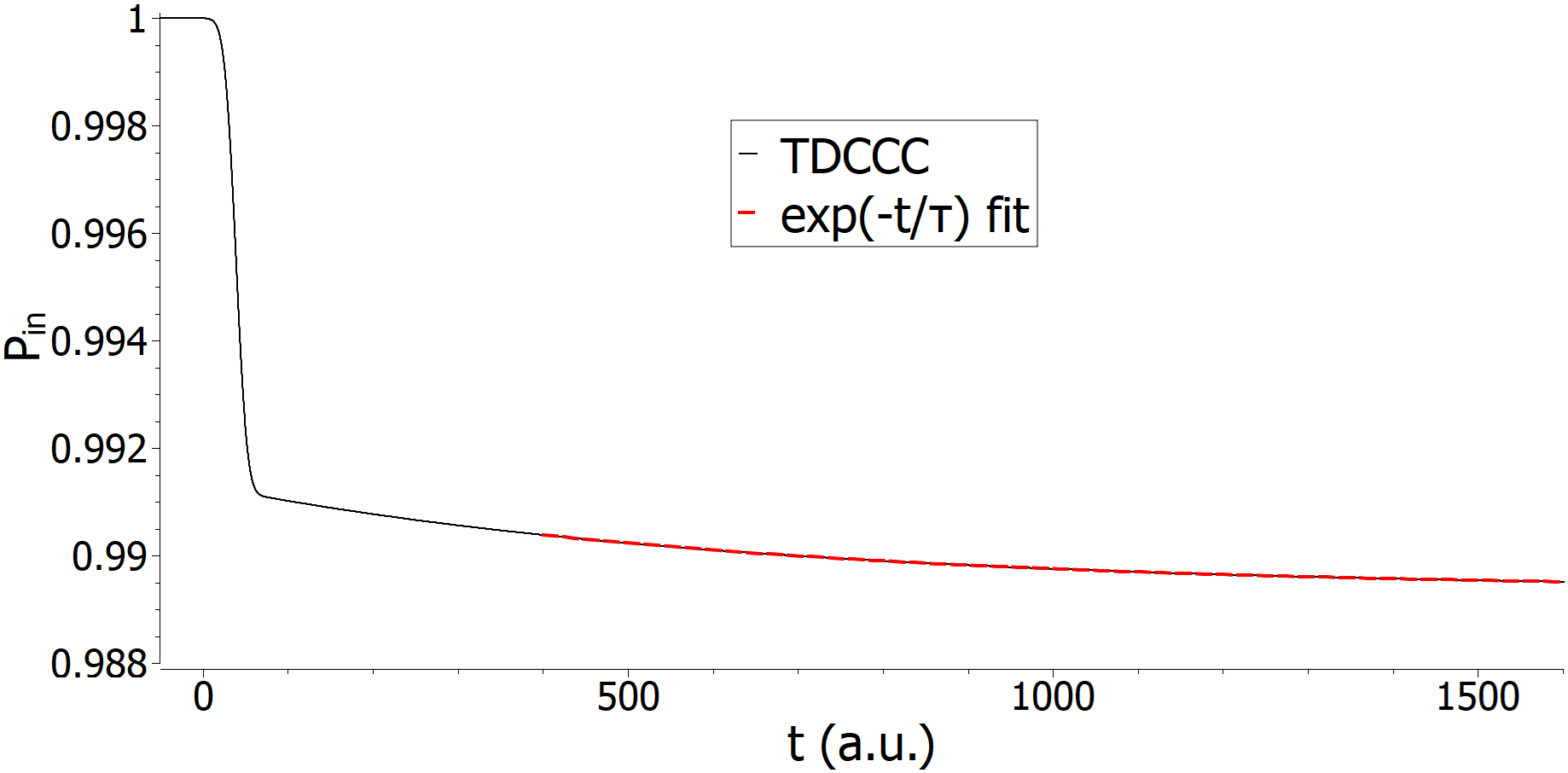}
	\includegraphics[width=0.7\textwidth]{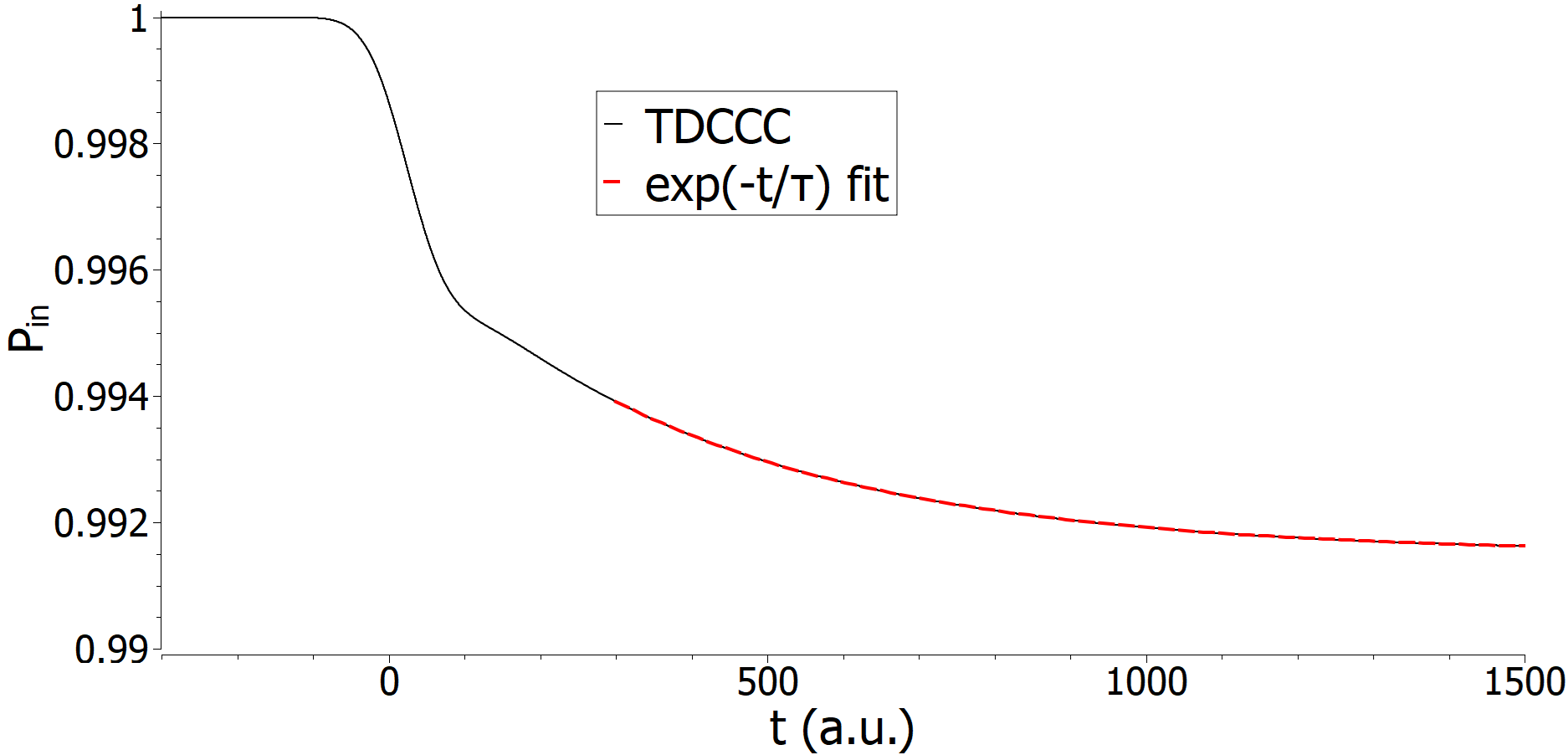}
	\caption{The probability $P_{in}$ of presence an electron in a
          radius $r<63$ a.u. around an ion after a XUV pulse as
          function of time: (top) He, $\omega_{XUV}=1.311$ a.u.,
          FWHM=0.6 fs; (bottom) Li$^+$, $\omega_{XUV}=5.427$ a.u.,
          FWHM=2.4 fs. The $t=0$ corresponds to the center of the XUV
          pulse.}
	\label{fig:Pin_t}
\end{figure}

\begin{table}[ht!]
\caption{Energy (in atomic units a.u.) and life-time (in atomic units
  a.u. and femto-seconds fs) of the lowest sp2$^+$ autoionizing states
  in He and Li$^+$.}
\label{tab:sp2}
\centering
\bt{|cccc|cccc|cc}
\hline
He sp2$^+$ & $E_e$ & \mc{2}{c|}{$\tau$} &Li$^+$ sp2$^+$ & $E_e$ & \mc{2}{c|}{$\tau$} \\
& a.u.& a.u. & fs && a.u. & a.u. & fs\\
\hline
& 1.311 & 618 & 15 && 2.747  & 396 & 9.6 \\
Expt \cite{Busto2018} & 1.307 & 703 & 17 &
Expt \cite{Carroll1977} & 2.808 & 363 & 8.7 \\
\hline
\et
\end{table}

In the second test, we calculate the photoelectron spectra driven by a
short broad-band XUV pulse. These spectra, displayed in
\Fref{fig:SpectrumShortPulse}, show clearly the sequence of spN$^+$,
$N\ge2$ autoionizing resonances when a multi-configuration expansion
with $N_s=5$ is included in \Eref{Psi_2e}. 
From these spectra we obtained the resonance energies $E_e$, which are shown in \Tref{tab:sp2}.

We also calculated the lifetimes of the meta-stable auto-ionizing states corresponding to the resonances. To do this, we calculated the time dependence of the probability of finding an electron inside the simulation boundary ($r<r_{ECS}=63$ a.u.) after exposure to a long XUV pulse precisely tuned to resonance. Such a pulse causes the population of the corresponding doubly excited state. After the leakage of electrons emitted due to direct ionization, the probability decreases due to the decay of the meta-stable state, as clear from \Fref{fig:Pin_t}. So, using an exponential approximation of the time dependence, one can obtain the lifetimes of the meta-stable state. The calculated lifetimes are shown in \Tref{tab:sp2} along with the state energies. They are in good agreement with the corresponding exact values shown in the same Table.

\acknowledgements{The author thanks Prof. A.~S. Kheifets for
  setting the problem for this work and for many stimulating
  discussions.  The work was supported by the Discovery Grant
  DP190101145 from the Australian Research Council.}

\end{document}